\documentclass[10pt,conference]{IEEEtran}
\newif\ifsubmission
\submissionfalse
\usepackage{cite}
\usepackage{amsmath,amssymb,amsfonts}
\usepackage{graphicx}
\usepackage{textcomp}
\usepackage[table]{xcolor}
\usepackage{booktabs}
\usepackage{adjustbox}
\usepackage{multirow}
\usepackage{listings}
\usepackage{algorithm}
\usepackage{algpseudocode}
\usepackage{microtype}
\usepackage[hidelinks]{hyperref}
\usepackage{pifont}
\usepackage[capitalise]{cleveref}
\usepackage[most]{tcolorbox}

\definecolor{tablemodelblue}{RGB}{247,250,255}
\definecolor{tablemodelgreen}{RGB}{248,252,248}
\definecolor{tablemodelgray}{RGB}{235,235,235}

\providecommand{\method}{\mbox{CoACT}}
\DeclareRobustCommand{\passatone}{\ifmmode\text{\normalfont\scshape pass@1}\else\textnormal{\textsc{pass@1}}\fi}
\newcommand{\bi}[1]{\textbf{\textit{#1}}}

\makeatletter
\newcommand{\tablecaptioninline}[2][]{%
  \refstepcounter{table}%
  \ifx\relax#1\relax
    \addcontentsline{lot}{table}{\protect\numberline{\thetable}#2}%
  \else
    \addcontentsline{lot}{table}{\protect\numberline{\thetable}#1}%
  \fi
  \noindent\textnormal{TABLE~\thetable: #2}\par\vspace{0.5em}%
}
\makeatother

\newcounter{rqanswer}
\crefname{rqanswer}{RQanswer}{RQanswers}
\Crefname{rqanswer}{RQanswer}{RQanswers}

\definecolor{answerorangeback}{RGB}{255,247,237}
\definecolor{answerorangeframe}{RGB}{245,158,11}
\definecolor{objectiveblueback}{RGB}{247,250,255}
\definecolor{objectiveblueframe}{RGB}{96,132,180}
\newtcolorbox[use counter=rqanswer]{rqanswer}[1][]{%
  enhanced,
  colback=answerorangeback,
  colframe=answerorangeframe!65,
  boxrule=0.3pt,
  arc=6pt,
  left=8pt, right=8pt, top=6pt, bottom=6pt,
  shadow={1pt}{-1pt}{0pt}{answerorangeframe!35},
  before upper={\textit{\ding{46}}~\textbf{Answer to RQ\therqanswer:}~},
  #1
}
\newtcolorbox{objectivebox}[1][]{%
  enhanced,
  colback=objectiveblueback,
  colframe=objectiveblueframe!70,
  boxrule=0.35pt,
  arc=3pt,
  left=0.5pt, right=5pt, top=-5pt, bottom=-1.5pt,
  before skip=4pt, after skip=4pt,
  #1
}

\newtcolorbox{objectivebox2}[1][]{%
  enhanced,
  colback=objectiveblueback,
  colframe=objectiveblueframe!70,
  boxrule=0.35pt,
  arc=3pt,
  left=0pt, right=5pt, top=1.5pt, bottom=-1pt,
  before skip=4pt, after skip=4pt,
  #1
}

\begin{document}

\title{CoACT: Action-Preserving Observation Compression for Coding Agents}

\ifsubmission\else
\author{
\IEEEauthorblockN{Haorui Chen}
\IEEEauthorblockA{\textit{College of AI} \\
\textit{Tsinghua University}\\
Beijing, China \\
hrchen@std.uestc.edu.cn}
\and
\IEEEauthorblockN{Yuancheng Zhu\textsuperscript{\ddag}}
\IEEEauthorblockA{\textit{School of Intelligence Science and Technology} \\
\textit{Nanjing University}\\
Nanjing, China \\
mileszhu@smail.nju.edu.cn}
\and
\IEEEauthorblockN{Yitong Zhang}
\IEEEauthorblockA{\textit{College of AI} \\
\textit{Tsinghua University}\\
Beijing, China \\
zhangyt42@buaa.edu.cn}
\and
\IEEEauthorblockN{Jia Li\textsuperscript{\dag}}
\IEEEauthorblockA{\textit{College of AI} \\
\textit{Tsinghua University}\\
Beijing, China \\
jia\_li@mail.tsinghua.edu.cn}
}
\fi

\maketitle

\ifsubmission\else
\begingroup
\renewcommand{\thefootnote}{}
\footnotetext[0]{\textsuperscript{\ddag}This work was done while Yuancheng Zhu was an intern at the College of AI,
Tsinghua University.
\par\textsuperscript{\dag}Corresponding author.}
\endgroup
\fi

\begin{abstract}
LLM-based coding agents solve software-engineering tasks through iterative interactions with development environments, where returned observations accumulate in the context and become a major source of inference cost.
Observation compression reduces this cost by shortening observations before they are appended to the context.
However, existing methods still exhibit an unsatisfactory efficiency-effectiveness trade-off, as they do not explicitly model how compression affects the agent's subsequent behavior.
This paper proposes \method{}, an action-preserving observation compression method for coding agents.
\method{} is built on next-action preservation (NAP), which requires a compressed observation to induce the same next action as the raw observation.
By checking the agent's immediate next action, NAP provides a practical signal for whether a compression preserves the information needed for continued task solving.
During training, a teacher model first generates multiple compressed candidates of each observation.
\method{} then uses an action-preservation reward based on NAP to filter out candidates that would change the agent's next action, and uses a length-reduction reward to choose compact candidates as supervision for a lightweight compressor.
Experiments on SWE-bench Verified with three agentic models show that \method{} reduces average total token consumption by 33.0\% while maintaining task-solving effectiveness close to the uncompressed agent.
\end{abstract}


\section{Introduction}
\label{sec:introduction}
 
LLM-based coding agents~\cite{yang2024swe, wang2025openhands, gao2025trae, xia2024agentless, cai2025ai} have shown strong potential in solving complex software engineering tasks.
Their strength comes from an interactive workflow with the development environment~\cite{yao2022react,zhang2026environmental}.
At each step, the coding agent proposes an action, receives an observation from the environment, and directly appends that observation to its context for subsequent steps.
As this process continues, observations accumulate rapidly and become a major source of inference cost, limiting the large-scale deployment of coding agents~\cite{xiao2025reducing, fan2025swe, stackoverflow2025, bai2026ai, gao2025more}.
For example, our analysis shows that observation tokens account for 45.7\% of total token consumption on SWE-bench Verified~\cite{jimenez2024swe} and up to 67.8\% on Terminal-Bench~\cite{merrill2026terminal}.

This observation-induced cost pressure has motivated a line of techniques (e.g., LongCodeZip~\cite{shi2025longcodezip} and LLMLingua-2~\cite{pan2024llmlingua}) that compress observations before they are appended to the agent context, which we refer to as \bi{observation compression}~\footnote{A detailed discussion of related work is provided in \Cref{sec:related}.}.
However, in practice, existing observation compression methods have not yet achieved a satisfactory \bi{efficiency-effectiveness trade-off}.
Some methods preserve task effectiveness but leave substantial efficiency gains unrealized, while others compress observations more aggressively but hurt task effectiveness or introduce extra recovery steps that offset the savings.
For example, on SWE-bench Verified, LLMLingua-2 compresses observations aggressively but drops \passatone{} from 57.0\% to 50.0\%, while LongCodeZip preserves task effectiveness but only reduces total token consumption from 3.795M to 3.165M.
This limitation raises one central question: \bi{how can observation compression better balance efficiency and effectiveness for coding agents?}

To answer this question, we first make the efficiency-effectiveness trade-off explicit by using total token consumption to represent efficiency and the widely adopted \passatone{} metric to measure effectiveness.
Given a compression strategy $C$ and the uncompressed baseline $\mathbb{I}$, observation compression can then be formulated as a \bi{constrained optimization problem}: minimize total token consumption while keeping the compressed agent within a small \passatone{} gap $\epsilon$ from the uncompressed agent:
\begin{objectivebox}
{\small
\begin{equation}
\min_C \mathrm{Tokens}(C)
\quad
\text{s.t.}\quad
\passatone{}(C)
\geq
\passatone{}(\mathbb{I}) - \epsilon .
\label{eq:intro_objective}
\end{equation}
}
\end{objectivebox}
\noindent This formulation highlights a key limitation of existing observation compression methods.
Most existing methods focus on shortening observations, but they do not quantitatively assess how compression affects the agent's task-solving effectiveness under the constraint in \Cref{eq:intro_objective}.
For example, LLMLingua-2 \cite{pan2024llmlingua} and SWE-Pruner~\cite{wang2026swe} use LLMs to compress observations, but lack an explicit constraint to control the impact of compression on effectiveness.
Thus, the compressor lacks explicit guidance for optimizing the trade-off between efficiency and effectiveness.
This missing guidance leads to two limitations in practice.
\ding{182} \emph{Effectiveness may be compromised.}
A compression candidate may remove information that is still needed by the agent, thereby reducing \passatone{}.
\ding{183} \emph{Efficiency gains may be limited.}
The compressor may also keep redundant content because it cannot quantitatively verify that a more compact candidate remains effective.
This motivates us to \bi{explicitly model the effectiveness constraint during observation compression}.

However, directly using \passatone{} as this effectiveness constraint is impractical, because it can only be checked at the end of an agent trajectory.
\ding{182} \emph{The constraint is too expensive to evaluate.}
When a single observation is compressed, the final task outcome is still unknown.
Directly checking whether this compression respects the \passatone{} constraint would require rolling out the remaining trajectory and evaluating the final task result.
Doing so is already expensive for one candidate and quickly becomes impractical across many observations and many possible compressions.
\ding{183} \emph{The constraint is too sparse to optimize.}
A single agent trajectory typically contains hundreds of observations, many of which may be compressed before the final outcome is observed.
The final success or failure therefore reflects the combined effect of many compression decisions.
It is therefore too sparse to provide reliable feedback for optimizing a particular compressed observation.
These difficulties show that \bi{the \passatone{} constraint in \Cref{eq:intro_objective} requires a more practical proxy}.

Fortunately, we find that the agent workflow itself points to such a proxy.
Although final task success is observed only at the end of the full trajectory, the effect of a compressed observation can appear immediately in \bi{the agent's next action}.
At step $t$, suppose the raw observation $o_t$ is replaced by a compressed observation $\hat{o}_t$, while the task goal and preceding trajectory remain unchanged.
If this replacement preserves the agent's next action, then the compressed agent takes the same next step as the uncompressed agent.
Naturally, if this holds at every compression step, the compressed agent follows the same action trajectory as the uncompressed agent and therefore achieves the same \passatone{}.
This makes next-action preservation a natural proxy for the \passatone{} constraint.
More fortunately, it also addresses the two difficulties above.
\ding{182} \emph{It is cheap enough to evaluate:} instead of rolling out the remaining trajectory, we only need to generate the agent's next action under $\hat{o}_t$ and compare it with the next action under $o_t$.
\ding{183} \emph{It is dense enough to optimize:} it provides a step-level feedback for each compressed observation, rather than a single feedback for all compressed observations in the trajectory.
We call this practical proxy \bi{next-action preservation (NAP)}.

Based on NAP, we rewrite \Cref{eq:intro_objective} into the following optimization problem, which is easier to solve in practice:
\begin{objectivebox2}
{\small
\begin{equation}
\min_C \mathrm{Tokens}(C)
\quad
\text{s.t.}
\quad
a_{t+1}^{C}=a_{t+1}^{\mathbb{I}},
\quad
\forall t .
\label{eq:nap_objective}
\end{equation}
}
\end{objectivebox2}
\noindent Here, $a_{t+1}^{\mathbb{I}}$ and $a_{t+1}^{C}$ denote the next actions under the raw and compressed observations, respectively.
We then propose \bi{\method{}}, a novel observation compression approach that trains a lightweight compressor to optimize observation compression under the NAP constraint.
Specifically, \method{} builds supervision from candidate compressions based on two complementary rewards.
\ding{182} The \emph{action-preservation reward} explicitly models the effectiveness constraint through NAP, keeping candidates that preserve the agent's original next action.
\ding{183} The \emph{length-reduction reward} explicitly models the efficiency objective by selecting more compact candidates under this constraint.
\method{} uses these reward-selected candidates to train the lightweight compressor through offline bootstrap and online alignment, enabling the deployed compressor to maximize length reduction for each new observation under the next-action preservation constraint.

To evaluate \method{}, we conduct experiments on SWE-bench Verified using three agentic models: Qwen3.5-35B-A3B~\cite{qwen3.5-35b-a3b_2026}, Deepseek-v4-Pro~\cite{xu2026deepseek}, and Gemini3-Flash~\cite{gemini_2025}.
Our results show that: \ding{182} \method{} reduces total token consumption across agentic models while keeping \passatone{} close to uncompressed settings, reducing total token consumption from 3.795M to 2.428M and improving \passatone{} from 57.0\% to 60.5\% on Qwen3.5-35B-A3B; \ding{183} compared with existing observation-compression baselines, \method{} achieves the best efficiency-effectiveness trade-off; \ding{184} \method{} can be combined with trajectory compression, reducing the total cost of AgentDiet~\cite{xiao2025reducing} from \$45.65 to \$25.88 while preserving task success; and \ding{185} ablation studies show that both the action-preservation reward and the length-reduction reward are necessary for optimizing the efficiency-effectiveness trade-off.

The contributions of this paper are summarized as follows.
\begin{itemize}
\item We formulate observation compression as a constrained optimization problem and introduce next-action preservation as a practical proxy for the effectiveness constraint.
\item We propose \method{} to solve this constrained optimization problem by training a lightweight compressor with reward-guided supervision that combines action preservation for effectiveness and length reduction for efficiency.
\item We evaluate \method{} on SWE-bench Verified, showing that it achieves a better efficiency-effectiveness trade-off than existing observation-compression baselines.
\end{itemize}

\section{Preliminaries and Related Work}
\label{sec:related}

In this section, we introduce the preliminaries needed to understand this paper and review related work.

\subsection{Preliminaries}
\label{sec:formalization}

LLM-based coding agents have shown strong ability in solving software-engineering tasks by interacting with development environments~\cite{yang2024swe, wang2025openhands, gao2025trae, xia2024agentless, li2026papers}.
Given a task goal $g$ (e.g., one issue), a coding agent repeatedly generates an action, receives an observation from the environment, and uses the accumulated trajectory to decide the next action~\cite{yao2022react}.
After step $t$, we denote the \bi{raw trajectory} available to the agent as follows:
\begin{equation}
  h_t = (a_1, o_1, a_2, o_2, \ldots, a_t, o_t),
\end{equation}
where $a_i$ is the agent action, such as a shell command or tool call, and $o_i$ is the observation returned by the environment, such as file content, search results, or command output.
The agent policy $\pi$ then samples the next action as
\begin{equation}
  a_{t+1} \sim \pi(\cdot \mid g, h_t).
\end{equation}

Because the full trajectory is fed back to the LLM context at each step, the cost of agent execution grows quickly as the trajectory expands.
More context tokens increase the computation of each subsequent LLM invocation.
Even when prefix KV-cache reuse is available, each newly appended token still needs to attend to the cached prefix, so the per-token computation grows roughly linearly with the current context length.
This cost pressure has motivated many compression methods for coding agents, which reduce execution cost by shortening the context fed to the LLM~\cite{xiao2025reducing, lindenbauer2025complexity, wang2026swe}.
We describe them through a compression function $C$:
\begin{equation}
  \tilde{h}_t = C(h_t), \quad
  a_{t+1} \sim \pi(\cdot \mid g, \tilde{h}_t),
\end{equation}
where $\tilde{h}_t$ is the \bi{compressed trajectory} seen by the agent.
When $\tilde{h}_t=h_t$, the agent receives the raw trajectory without any compression.
We denote this uncompressed strategy as $\mathbb{I}$.

According to where $C$ is applied, compression methods can be divided into two orthogonal families.
\ding{182} \emph{Trajectory compression} applies $C$ to the whole accumulated trajectory and decides which historical content should remain visible at step $t$.
\ding{183} \emph{Observation compression} applies $C$ only to the newest observation before it is appended to the context.
The latter has recently attracted increasing attention because observations are a major source of token consumption, and compressing the newest observation before it is appended to the context can preserve valuable KV-cache reuse.
Therefore, this paper mainly focuses on observation compression.

\subsection{Trajectory Compression}
\label{sec:trajectory_compression}

Trajectory compression reduces inference cost by directly rewriting the accumulated trajectory.
At step $t$, such methods construct the compressed trajectory as
\begin{equation}
  \tilde{h}_t = C_t(a_1, o_1, \ldots, a_t, o_t),
\end{equation}
where $C_t$ may delete, truncate, summarize, or reorganize any content from previous steps.
For example, a sliding-window strategy~\cite{lindenbauer2025complexity} keeps only the most recent $W$ steps:
\begin{equation}
  \tilde{h}_t =
  (a_{t-W+1}, o_{t-W+1}, \ldots, a_t, o_t).
\end{equation}

Both industrial systems and academic studies follow this direction~\cite{sun2025scaling, ye2025agentfold, lu2025scaling, wan2025compass, kang2025acon, feng2026agentswing, liu2026context, verma2026active}.
In industrial systems, Claude Code~\cite{claude_code} and Cursor~\cite{cursorCursor} summarize the context when it becomes large.
In academic studies, Trae Agent~\cite{gao2025trae} and mini SWE-agent~\cite{yang2024swe} truncate long trajectory content to a fixed budget, while Lindenbauer et al.~\cite{lindenbauer2025complexity} delete old trajectory content after a fixed delay.
AgentDiet~\cite{xiao2025reducing} further uses an LLM module to identify useless, redundant, or expired historical content and removes it within a sliding window.

These methods reduce tokens by rewriting trajectory that has already entered the context.
Therefore, the trajectory at step $t$ is not obtained by simply appending the new action and observation to the trajectory at step $t-1$.
Formally, in general,
\begin{equation}
  \tilde{h}_t \neq (\tilde{h}_{t-1}, a_t, o_t).
\end{equation}
This rewriting exposes an inherent limitation of trajectory compression: its token savings come with the cost of invalidating prefix KV-cache reuse.
Once earlier tokens are compressed, the cached key-value states computed for the previous prefix no longer match the current context.
The model must therefore prefill the rewritten context again, which requires full self-attention over $\tilde{h}_t$ and incurs $O(|\tilde{h}_t|^2)$ computation.
Thus, the token savings from trajectory compression can be partly offset by recomputation over rewritten contexts.
This motivates new compression methods that reduce total token consumption without disrupting valuable KV-cache reuse.

\subsection{Observation Compression}
\label{sec:observation_compression}

Given the inherent limitation of trajectory compression, recent work~\cite{ren2026self, jia2026compressing} has turned to observation compression.
Unlike trajectory compression, observation compression leaves the historical context unchanged and directly compresses only the new observation before it is appended to the context.
Formally, given the previous compressed trajectory $\tilde{h}_{t-1}$, the current action $a_t$, and the raw observation $o_t$, observation compression produces new compressed trajectory as
\begin{equation}
  \tilde{h}_t
  =
  (\tilde{h}_{t-1}, a_t, \hat{o}_t)
  =
  C(\tilde{h}_{t-1}, a_t, o_t).
\end{equation}
Because the historical context is unchanged, the context at step $t$ extends $\tilde{h}_{t-1}$ with only the new action and compressed observation.
This naturally preserves valuable KV-cache reuse.

Existing observation compression methods mainly differ in how they decide which parts of a new observation should be kept.
General text-compression methods, such as LLMLingua~\cite{jiang2023llmlingua}, LongLLMLingua~\cite{jiang2024longllmlingua}, and LLMLingua-2~\cite{pan2024llmlingua}, estimate token importance and remove tokens with low importance scores.
Provence~\cite{chirkova2025provence} and CPC~\cite{liskavets2025prompt} also follow this direction by training encoder-based compressors.
Code-oriented compression methods~\cite{he2025codepromptzip, jia2026compressing} introduce programming-language structure into compression.
For example, LongCodeZip~\cite{shi2025longcodezip} groups code into chunks and prunes chunks with entropy-guided scores.
SWE-Pruner~\cite{wang2026swe} further uses the agent's current focus: it first asks the agent what it is looking for, and then uses this description to keep observation lines that appear relevant.

However, in practice, we find that existing observation compression methods have not yet achieved a satisfactory efficiency-effectiveness trade-off.
% Some methods preserve task effectiveness but leave substantial efficiency gains unrealized, while others compress observations more aggressively but hurt task effectiveness or introduce recovery steps that offset the savings.
On SWE-bench Verified, for example, LLMLingua-2 achieves effective observation compression but drops \passatone{} from 57.0\% to 50.0\%.
LongCodeZip preserves task effectiveness but only reduces total token consumption from 3.795M to 3.165M per instance.
These limitations call for observation compression methods that better balance efficiency and effectiveness for coding agents.

\section{Methodology}
\label{sec:methodology}

In this section, we propose \method{}, an observation compression approach designed to improve the efficiency-effectiveness trade-off for coding agents.

\begin{figure*}[ht]
  \centering
  \includegraphics[width=0.9\linewidth]{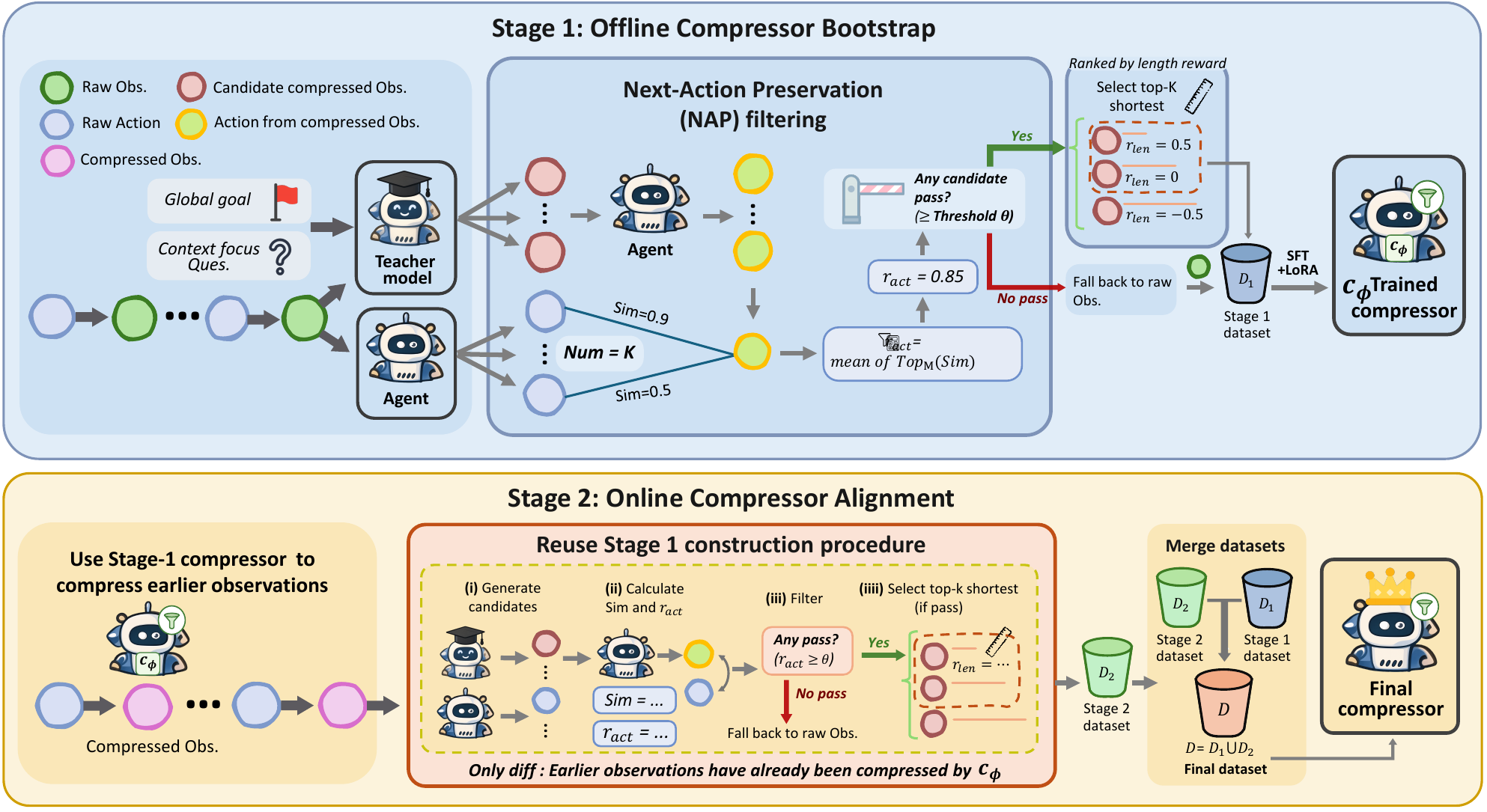}
  \caption{Overview of \method{}.}
  \label{fig:overview}
  \vspace{-10pt}
\end{figure*}

\subsection{Overview}
\label{sec:method_overview}

Observation compression should reduce total token consumption while explicitly modeling the effectiveness constraint.
\Cref{eq:intro_objective} captures this goal as a constrained optimization problem, with total token consumption as the efficiency objective and \passatone{} as the effectiveness constraint.
This formulation makes the desired trade-off explicit, but it is difficult to solve directly because \passatone{} can be measured only after the full trajectory finishes.

As discussed in \Cref{sec:introduction}, \method{} addresses this difficulty by using next-action preservation (NAP) as the proxy for the \passatone{} constraint.
Replacing the \passatone{} constraint in \Cref{eq:intro_objective} gives \Cref{eq:nap_objective}.
To solve \Cref{eq:nap_objective}, \method{} trains a lightweight compressor with reward-selected supervision.
\ding{182} For each raw observation, a teacher model first generates multiple candidate compressions.
\ding{183} \method{} then selects training candidates through reward-guided selection (\Cref{sec:supervision_construction}): the action-preservation reward keeps candidates that satisfy the NAP constraint, and the length-reduction reward prefers shorter candidates among them.
\ding{184} The selected candidates are used to train the compressor through offline bootstrap and online alignment (\Cref{sec:compressor_training}).
\ding{185} During deployment, only the trained compressor is placed in the agent workflow to compress each new observation (\Cref{sec:deployment}).

\subsection{Reward-Guided Supervision Construction}
\label{sec:supervision_construction}

\Cref{eq:nap_objective} specifies which compressed observations are desirable, but it does not provide direct supervision for training a compressor.
\method{} therefore constructs supervision from teacher-generated candidate compressions by first enforcing the NAP constraint and then favoring shorter candidates.

At step $t$, let $\bar{h}_{t-1}$ denote the trajectory before the current action-observation pair $(a_t,o_t)$ is appended.
A teacher model $T$ receives the task goal $g$, the current trajectory $\bar{h}_{t-1}$, the current action $a_t$, and the raw observation $o_t$, and generates $N$ candidate compressions independently:
\begin{equation}
\{\hat{o}_t^{(j)}\}_{j=1}^{N}
=
T(g,\bar{h}_{t-1},a_t,o_t).
\end{equation}
Because these candidates may differ in how well they satisfy \Cref{eq:nap_objective}, \method{} uses the following rewards to select candidates aligned with this formulation.

\vspace{2.5pt}
\bi{Action-Preservation Reward.}
The action-preservation reward checks whether a candidate satisfies the NAP constraint.
The constraint in \Cref{eq:nap_objective} compares the next action under the compressed observation with the next action under the raw observation.
In practice, an LLM-based agent may produce several valid next actions under stochastic decoding, so using only one action as the reference would be brittle.
\method{} therefore independently samples $K$ reference actions under the raw observation:
\begin{equation}
a_{t+1}^{(k)} \sim \pi(\cdot \mid g,\bar{h}_{t-1},a_t,o_t),
\quad
k=1,\ldots,K .
\end{equation}

For a candidate compression $\hat{o}_t^{(j)}$, \method{} replaces only the raw observation $o_t$ and keeps the task goal, trajectory, and current action unchanged.
The agent then produces the next action under this compressed observation with greedy decoding:
\begin{equation}
\hat{a}_{t+1}^{(j)}
=
\arg\max_a
\pi(a\mid g,\bar{h}_{t-1},a_t,\hat{o}_t^{(j)}).
\end{equation}

Let $\mathrm{sim}(\cdot,\cdot)\in[0,1]$ denote a given action-similarity function, such as text-similarity matching, where a higher value indicates that two actions express more similar operational intent.
Given a hyperparameter $M\leq K$, \method{} compares the action under the compressed observation with all reference actions and computes the action-preservation reward as the mean of the top $M$ similarity scores:
\begin{equation}
\begin{aligned}
\mathcal{V}_t^{(j)}
&=
\left\{
\mathrm{sim}\!\left(
\hat{a}_{t+1}^{(j)},
a_{t+1}^{(k)}
\right)
\right\}_{k=1}^{K},
\\
r_{\mathrm{act}}\!\left(\hat{o}_t^{(j)}\right)
&=
\mathrm{mean}\!\left(
\mathrm{Top}_{M}
\left(\mathcal{V}_t^{(j)}\right)
\right).
\end{aligned}
\label{eq:action_reward}
\end{equation}
\method{} then keeps candidates whose action-preservation reward reaches a threshold $\theta$:
\begin{equation}
P_t
=
\left\{
\hat{o}_t^{(j)}
\mid
r_{\mathrm{act}}\!\left(\hat{o}_t^{(j)}\right) \geq \theta
\right\}.
\label{eq:accepted_candidates}
\end{equation}

\vspace{2.5pt}
\bi{Length-Reduction Reward.}
After $P_t$ is formed, the remaining candidates have already passed the NAP constraint.
The length-reduction reward then optimizes the token consumption objective.
That is, \method{} prefers shorter candidates only among candidates whose next actions remain aligned with the raw observation.

For any candidate $\hat{o}_t^{(j)}\in P_t$, \method{} defines length-reduction reward as the number of tokens removed from the raw observation:
\begin{equation}
r_{\mathrm{len}}(\hat{o}_t^{(j)})
=
|o_t|-|\hat{o}_t^{(j)}|.
\label{eq:length_reward}
\end{equation}
Candidates with larger token savings receive higher length-reduction rewards, so \method{} selects the top-$k$ candidates in $P_t$ by $r_{\mathrm{len}}$ for subsequent training.

In some cases, $P_t$ is empty because no candidate compression reaches the action-preservation threshold.
We treat this case as indicating that all content in the observation may need to be kept to preserve the next action.
\method{} then uses the raw observation $o_t$ for later training.
This fallback teaches the compressor to leave such observations unchanged.

\subsection{Compressor Training}
\label{sec:compressor_training}

\Cref{sec:supervision_construction} discusses how to construct supervision data aligned with \Cref{eq:nap_objective}.
In this subsection, we describe how \method{} uses this data to train the compressor.
The training process consists of two stages: offline bootstrap and online alignment.
The offline bootstrap stage constructs supervision under raw trajectories and trains a cold-start compressor, while the online alignment stage constructs supervision under trajectories whose earlier observations have already been compressed and further aligns the compressor.
For clarity, \Cref{alg:training} summarizes the training procedure.

\vspace{2.5pt}
\bi{Offline Bootstrap.}
This stage uses trajectories collected from the uncompressed agent.
At each step $t$, $\bar{h}_{t-1}$ is instantiated as the raw trajectory $h_{t-1}$, and the current action $a_t$ and observation $o_t$ are also from the uncompressed trajectory.
\method{} applies reward-guided supervision construction to $(g,\bar{h}_{t-1},a_t,o_t)$, producing the offline training set $\mathcal{D}_1$.
Supervised fine-tuning on $\mathcal{D}_1$ gives the cold-start compressor $c_{\phi_1}$.

\vspace{2.5pt}
\bi{Online Alignment.}
The offline stage computes rewards under raw trajectories.
This creates a mismatch with deployment, where earlier observations in the trajectory have already been compressed by the compressor.
To reduce this mismatch, the online alignment stage runs the agent with $c_{\phi_1}$.
At each step $t$, $\bar{h}_{t-1}$ is instantiated as the compressed trajectory $\tilde{h}_{t-1}$, which already contains earlier observations compressed by $c_{\phi_1}$.
\method{} repeats reward-guided supervision construction on $(g,\tilde{h}_{t-1},a_t,o_t)$, producing the online training set $\mathcal{D}_2$.
\method{} then continues training $c_{\phi_1}$ on $\mathcal{D}_2$ to obtain $c_{\phi_2}$, which serves as the final compressor.

\begingroup
\begin{algorithm}[t]
\caption{\method{} Training}
\label{alg:training}
\small
\begin{algorithmic}[1]
\Require Agent policy $\pi$, teacher model $T$, training tasks
\Ensure Final compressor $c_{\phi_2}$

\State Collect raw trajectory $h_{t-1}$, actions $a_t$, and observations $o_t$
\State Construct $\mathcal{D}_1$ by reward-guided supervision construction under $(g,h_{t-1},a_t,o_t)$
\State Train the cold-start compressor $c_{\phi_1}$ on $\mathcal{D}_1$
\State Run the agent with $c_{\phi_1}$ to collect compressed trajectories $\tilde{h}_{t-1}$
\State Construct $\mathcal{D}_2$ by reward-guided supervision construction under $(g,\tilde{h}_{t-1},a_t,o_t)$
\State Continue training $c_{\phi_1}$ on $\mathcal{D}_2$ to obtain $c_{\phi_2}$
\State \Return $c_{\phi_2}$
\end{algorithmic}
\end{algorithm}
\endgroup

\subsection{Deployment}
\label{sec:deployment}

After training, \method{} deploys only the final lightweight compressor $c_{\phi_2}$ inside the agent workflow. Specifically, at step $t$, the agent produces an action $a_t$ and receives the raw observation $o_t$ from the environment.
The compressor produces:
\begin{equation}
\hat{o}_t = c_{\phi_2}(g,\tilde{h}_{t-1},a_t,o_t).
\end{equation}
\method{} then appends $\hat{o}_t$ rather than $o_t$ to the trajectory, and the agent proceeds to the next step.
Because \method{} compresses only the newly returned observation before it enters the context, the historical trajectory remains unchanged and valuable KV cache is preserved.

\section{Experimental Setup}
\label{sec:eval_setup}

To assess \method{}, we conduct comprehensive experiments to answer five Research Questions (RQs).
In this section, we present the details of our experimental setup.

\subsection{Research Questions}
\label{sec:rq}

\emph{RQ1: Can \method{} reduce inference consumption while preserving \passatone{} across agentic models?}
To answer it, we compare \method{} with Vanilla using Qwen3.5-35B-A3B~\cite{qwen3.5-35b-a3b_2026}, Deepseek-v4-Pro~\cite{xu2026deepseek}, and Gemini3-Flash~\cite{gemini_2025}.

\emph{RQ2: Can \method{} achieve a better efficiency-effectiveness trade-off than existing observation compression methods?}
To answer it, we compare \method{} with three representative baselines, including LLMLingua-2~\cite{pan2024llmlingua}, LongCodeZip~\cite{shi2025longcodezip}, and SWE-Pruner~\cite{wang2026swe}, using Qwen3.5-35B-A3B and Deepseek-v4-Pro.

\emph{RQ3: Can \method{} complement trajectory compression for further cost reduction?}
This RQ examines whether \method{} can be combined with complementary trajectory compression methods.
To answer it, we insert \method{} into two trajectory compression baselines and compare the combined methods against the baselines alone~\cite{lindenbauer2025complexity, xiao2025reducing}.

\emph{RQ4: How do the two rewards contribute to the trained compressor?}
This RQ examines whether the action-preservation reward maintains next-action preservation and whether the length-reduction reward drives token reduction.
To answer it, we compare the \method{} with compressors trained without the action-preservation reward and compressors trained without the length-reduction reward.

\emph{RQ5: Does online compressor alignment improve the compressor beyond offline bootstrap?}
To answer it, we compare the compressor trained only with offline supervision and the final compressor trained with both offline and online supervision.

\subsection{Models \& Scaffolds}
\label{sec:model}

In this subsection, we introduce the agentic models used for evaluation, the model used to initialize the observation compressor, and the agent scaffold used in our experiments.
\ding{182} \bi{Agentic models.}
We evaluate \method{} with three popular agentic models: Qwen3.5-35B-A3B, Deepseek-v4-Pro, and Gemini3-Flash. 
\ding{183} \bi{Observation compressor.}
Following prior work~\cite{wang2026swe}, we initialize the observation compressor with Qwen3.5-4B~\cite{qwen3.5-4b_2026} in all experiments.
\ding{184} \bi{Agent scaffold.}
Following prior work~\cite{wang2026swe}, we build the agent with widely used mini-swe-agent~\cite{yang2024swe}.
This scaffold provides only a bash interface to the agent: the agent interacts with the repository by issuing bash commands and receiving command outputs.

\begin{table*}[t]
    \centering
    \tablecaptioninline[RQ1 results across agentic models.]{
    Comparison between Vanilla and \method{} across three agentic models.
    \textbf{Bold} values indicate the better result for each model.
    }
    \label{tab:rq1_cross_model}
    \begingroup
    \footnotesize
    \setlength{\tabcolsep}{9pt}
    \renewcommand{\arraystretch}{1.12}
    \begin{adjustbox}{max width=0.98\textwidth,center}
    \begin{tabular}{clccccccc}
    \toprule
    Model & Method & \passatone{} $\uparrow$ & Steps $\downarrow$ & S-Steps $\downarrow$ &
    C-In $\downarrow$ & U-In $\downarrow$ & Out $\downarrow$ &
    Tokens $\downarrow$ \\
    \midrule
    \rowcolor{tablemodelblue}
    & Vanilla
    & 57.0 & 92.6 & 75.0 & 3.45 & 0.32 & 0.024
    & 3.795 \\
    \rowcolor{tablemodelblue}
    \multirow{-2}{*}{\textbf{Qwen3.5-35B-A3B}} & \quad+\method{}
    & \textbf{60.5} & \textbf{77.5} & \textbf{66.2}
    & \textbf{2.12} & \textbf{0.24} & \textbf{0.020}
    & \textbf{2.428} \\
    \midrule
    \rowcolor{tablemodelgreen}
    & Vanilla
    & \textbf{76.5} & \textbf{37.7} & \textbf{33.6}
    & 1.03 & 0.025 & 0.013
    & 1.068 \\
    \rowcolor{tablemodelgreen}
    \multirow{-2}{*}{\textbf{Deepseek-v4-Pro}} & \quad+\method{}
    & 75.0 & 38.1 & 36.6
    & \textbf{0.82} & \textbf{0.015} & 0.013
    & \textbf{0.868} \\
    \midrule
    \rowcolor{tablemodelgray}
    & Vanilla
    & 73.5 & 39.3 & \textbf{35.2}
    & 0.00 & 0.677 & 0.009
    & 0.685 \\
    \rowcolor{tablemodelgray}
    \multirow{-2}{*}{\textbf{Gemini3-Flash}} & \quad+\method{}
    & \textbf{77.5} & \textbf{38.0} & 37.4
    & 0.00 & \textbf{0.395} & \textbf{0.008}
    & \textbf{0.422} \\
    \midrule
    & Vanilla
    & 69.0 & 56.5 & 47.9
    & 1.49 & 0.341 & 0.015
    & 1.849 \\
    & \quad+\method{}
    & \textbf{71.0} & \textbf{51.2} & \textbf{46.7}
    & \textbf{0.98} & \textbf{0.217} & \textbf{0.014}
    & \textbf{1.239} \\
    \multirow{-3}{*}{\textbf{Average}} & Change
    & +2.0 pp & -9.4\% & -2.5\%
    & -34.2\% & -36.4\% & -6.7\%
    & -33.0\% \\
    \bottomrule
    \end{tabular}
    \end{adjustbox}
    \endgroup
    \vspace{-8pt}
\end{table*}

\subsection{Datasets \& Benchmarks}
\label{sec:bench}

In this subsection, we introduce the data used to train the observation compressor and the benchmark used in evaluation.

\bi{Training data.}
\method{} requires compressor training data, so we construct model-specific training sets from SWE-smith~\cite{yang2026swe} to avoid overlap with the evaluation benchmark.
For each target agentic model, we collect trajectories with that model and follow the pipeline described in \Cref{sec:compressor_training}.
Here we detail the Qwen3.5-35B-A3B setting: we randomly sample 50 SWE-smith instances and collect trajectories by running the uncompressed Qwen3.5-35B-A3B agent with the mini-swe-agent scaffold.
The offline bootstrap stage uses 686 eligible observations and yields $\mathcal{D}_1$ with 2,204 examples.
The online alignment stage uses 839 eligible observations and yields $\mathcal{D}_2$ with 2,662 examples.
The final Qwen training set $\mathcal{D}_1\cup\mathcal{D}_2$ contains 4,866 examples.
The Deepseek-v4-Pro and Gemini3-Flash training sets are constructed with the same pipeline and have comparable scale, with detailed statistics provided in the \textit{Supplementary Material}.

\bi{Evaluation benchmark.}
We evaluate \method{} on SWE-bench Verified~\cite{jimenez2024swe}.
Following prior work~\cite{xiao2025reducing}, we randomly sample 200 instances for evaluation and run all compared baselines on the same instances.

\subsection{Baselines}
\label{sec:baseline}

We include three types of baselines to evaluate \method{}.
\emph{Vanilla} measures the behavior of the uncompressed agent.
\emph{Observation compression baselines} evaluate whether \method{} achieves a better efficiency-effectiveness trade-off than existing observation compression methods.
\emph{Trajectory compression baselines} evaluate whether \method{} can be combined with orthogonal trajectory compression for further cost reduction.

\bi{Uncompressed baseline.}
\emph{Vanilla} runs the agent with the raw trajectory and serves as the uncompressed reference.

\bi{Observation compression baselines.}
\ding{182} \emph{LLMLingua-2}~\cite{pan2024llmlingua} is a general text compressor.
It assigns each token a keep-or-drop score and keeps the highest-scored tokens until the target compression ratio is reached.
We include it as a representative general-purpose text compression method.
\ding{183} \emph{LongCodeZip}~\cite{shi2025longcodezip} is a code-aware compressor.
It splits code observations into chunks, scores these chunks, and keeps the chunks judged useful under a token budget.
We include it as a representative compression method that uses code structure.
\ding{184} \emph{SWE-Pruner}~\cite{wang2026swe} is designed for coding agents.
It first asks the agent to describe what it is looking for, then uses this hint to score observation lines and keep the lines judged useful.
We include it as the closest baseline because it is also designed for coding agents.

\bi{Trajectory compression baselines.}
\ding{182} \emph{Sliding Window}~\cite{lindenbauer2025complexity} compresses the accumulated trajectory by keeping only the most recent action-observation steps and dropping earlier history.
We include it as the simplest trajectory-compression baseline.
\ding{183} \emph{AgentDiet}~\cite{xiao2025reducing} compresses the accumulated trajectory more selectively.
It asks an LLM module to inspect recent trajectory steps and remove content judged redundant or expired.
We include it as a more adaptive trajectory-compression baseline.

\subsection{Metrics}
\label{sec:eval_metrics}

We evaluate the efficiency-effectiveness trade-off of \method{} with two primary groups of metrics: efficiency metrics and effectiveness metrics.
% We additionally use NAP-related metrics to measure how well compression preserves the agent's next action.

\bi{Efficiency metrics.}
\ding{182} \emph{Cached Input Tokens (C-In, million tokens per instance)} and \ding{183} \emph{Uncached Input Tokens (U-In, million tokens per instance)} denote agent input tokens processed with and without prefix KV-cache reuse, respectively.
\ding{184} \emph{Output Tokens (Out, million tokens per instance)} denotes tokens generated by the agentic model.
\ding{185} \emph{Compressor Tokens (Comp, million tokens per instance)} denotes tokens consumed by the compression model.
\ding{186} \emph{Total Tokens (Tokens, million tokens per instance)} is the most direct efficiency metric and denotes the sum of C-In, U-In, Out, and Comp.
\ding{187} \emph{Interaction Steps (Steps, steps per instance)} measures the average number of interaction steps over all instances.
\ding{188} \emph{Solved Interaction Steps (S-Steps, steps per solved instance)} measures the average number of interaction steps over solved instances.
Together, Steps and S-Steps show whether compression causes additional tool calls and affects execution efficiency.

\bi{Effectiveness metrics.}
\emph{Task Success (\passatone{}, \%)} is the most direct effectiveness metric and measures the percentage of tasks solved by the agent.
An instance is solved if the generated patch passes the corresponding tests.

\subsection{Other Implementation Details}
\label{sec:implementation}

Following the official model inference recommendations~\cite{qwen3.5-35b-a3b_2026}, we set the temperature to 0.6 and top-$p$ to 0.95 for inference in all experiments unless otherwise stated.
For all baselines, we keep common hyperparameters the same as \method{} and follow the original papers for method-specific settings.

For \method{}, Gemini3-Flash serves as the teacher model and generates $N=8$ candidate compressions for each observation.
For reward-guided selection, the number of reference actions is $K=8$, the top $M$ action-similarity scores are aggregated with $M=3$, the action-preservation threshold is $\theta=0.6$, and the number of length-selected candidates is $k=4$.
The selected examples are used for supervised fine-tuning with LoRA~\cite{hu2022lora}, where we set the rank to 64 and alpha to 128.
Inspired by prior work on command similarity measurement~\cite{hussain2021command}, our action-similarity function parses each action and extracts operation fields such as target file, search pattern, and viewed line range.
We then compute a normalized similarity by matching operation types and comparing their associated fields.
To facilitate reproducibility, we provide additional implementation details in \textit{Supplementary Materials}.

\section{Experimental Results}
\label{sec:eval_results} 

\subsection{RQ1: Can \method{} reduce inference consumption while preserving \passatone{} across agentic models?}
In this RQ, we examine whether \method{} consistently reduces inference consumption while keeping \passatone{} close to Vanilla across different agentic models.

\vspace{1pt} \bi{Setting.}
We apply \method{} and Vanilla separately to Qwen3.5-35B-A3B, Deepseek-v4-Pro, and Gemini3-Flash.
For each agentic model, we evaluate both methods on the same SWE-bench Verified instances with the same mini-swe-agent scaffold, using the metrics defined in \Cref{sec:eval_metrics}.

\begin{figure}[t]
    \centering
    \includegraphics[width=0.9\columnwidth]{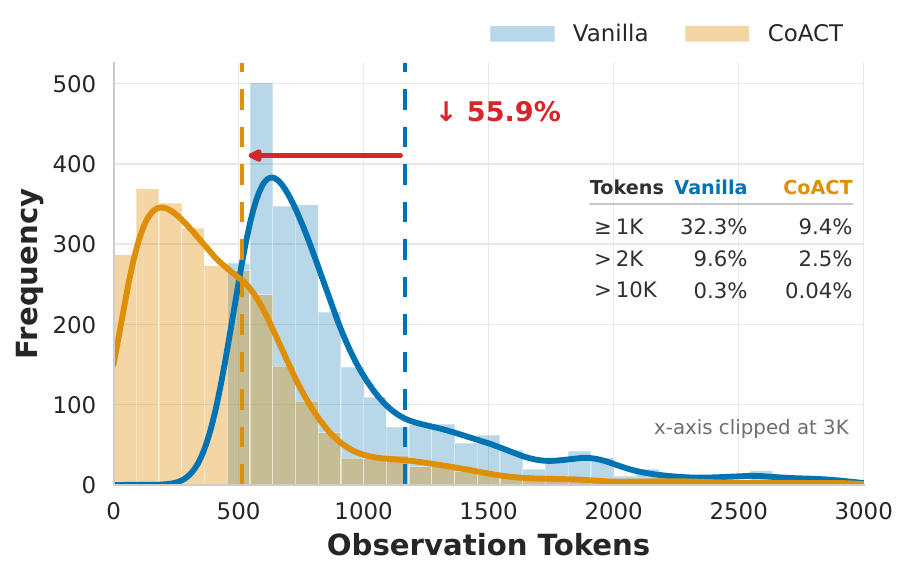}
    \caption{
    Observation-token distributions of Vanilla and \method{} under Qwen3.5-35B-A3B.
    Bars show frequency, and curves show smoothed distributions.
    }
    \label{fig:rq1_observation_tokens}
    \vspace{-10pt}
\end{figure}
    
\vspace{1pt} \bi{Results.}
\Cref{tab:rq1_cross_model} reports the effectiveness and efficiency metrics of Vanilla and \method{} across the three agentic models.

\textbf{\method{} consistently improves efficiency while preserving effectiveness across agentic models.}
As shown in \Cref{tab:rq1_cross_model}, \method{} improves efficiency for all three agentic models.
For Qwen3.5-35B-A3B, it reduces total token consumption from 3.795M to 2.428M.
For Deepseek-v4-Pro, it reduces total token consumption from 1.068M to 0.868M.
For Gemini3-Flash, it reduces total token consumption from 0.685M to 0.422M.
Meanwhile, \method{} keeps \passatone{} close to Vanilla.
On Deepseek-v4-Pro, for example, \passatone{} decreases only from 76.5\% to 75.0\%.
More encouragingly, \method{} can also bring modest effectiveness gains.
It improves \passatone{} on Qwen3.5-35B-A3B and Gemini3-Flash by 3.5 and 4.0 percentage points, respectively.
We attribute these gains to the fact that raw observations often contain irrelevant content that can distract the agent's reasoning, while \method{} removes such content before it enters the context.

\textbf{The efficiency gain comes from directly shortening observations, especially long observations.}
The results in \Cref{tab:rq1_cross_model} show that \method{} reduces inference cost, but they do not show where the savings come from.
Since \method{} acts on observations before they enter the trajectory, we further analyze whether it actually shortens the observations appended to the agent context.
Specifically, we use the Qwen3.5-35B-A3B runs in RQ1 and collect the observation length at each interaction step.
For Vanilla, we record the length of the raw observation.
For \method{}, we record the length of the compressed observation appended after compression.
\Cref{fig:rq1_observation_tokens} plots the resulting distributions.
Vanilla has a clear long-observation tail: 32.3\% of its observations contain more than 1K tokens, including 9.6\% with more than 2K tokens.
After applying \method{}, this tail becomes much thinner. For example, the proportion of observations exceeding 2K tokens decreases from 9.6\% to 2.5\%.
This shift suggests that shortening long observations before they are appended to the trajectory partly explains the cost savings observed in \Cref{tab:rq1_cross_model}.

\begin{rqanswer}
\method{} improves efficiency across all evaluated agentic models while keeping \passatone{} close to, and sometimes higher than, Vanilla.
\end{rqanswer}

\begin{table*}[t]
    \centering
    \tablecaptioninline[RQ2 observation-compression baseline results.]{
    Comparison with observation compression baselines.
    \textbf{Bold} and \underline{underlined} values mark the best and second-best results for each model.
    }
    \label{tab:rq2_observation_baselines}
    % \begingroup
    % \footnotesize
    \newcommand{\chg}[1]{\textsuperscript{\scriptsize #1}}
    \setlength{\tabcolsep}{7pt}
    \renewcommand{\arraystretch}{0.95}
    \begin{adjustbox}{max width=1\textwidth,center}
    \begin{tabular}{@{}lcccccccc@{}}
    \toprule
    Method & \passatone{} $\uparrow$ & Steps $\downarrow$ & S-Steps $\downarrow$ &
    C-In $\downarrow$ & U-In $\downarrow$ & Out $\downarrow$ &
    Comp $\downarrow$ & Tokens $\downarrow$ \\
    \midrule
    \rowcolor{tablemodelblue}
    \multicolumn{9}{c}{\textbf{Qwen3.5-35B-A3B}} \\
    \midrule
    Vanilla
    & 57.0 & 92.6 & 75.0 & 3.45 & 0.32 & 0.024
    & \textbf{0.000} & 3.795 \\
    LLMLingua-2
    & 50.0\chg{-7.0} & 195.0\chg{+111\%} & 170.2\chg{+127\%}
    & 9.45\chg{+174\%} & 0.77\chg{+141\%} & 0.034\chg{+42\%}
    & 0.078\chg{+0.078} & 10.329\chg{+172\%} \\
    LongCodeZip
    & \underline{60.0}\chg{+3.0} & \underline{86.4}\chg{-7\%} & \underline{68.8}\chg{-8\%}
    & \underline{2.82}\chg{-18\%} & \underline{0.29}\chg{-9\%} & \underline{0.022}\chg{-8\%}
    & \underline{0.038}\chg{+0.038} & \underline{3.165}\chg{-17\%} \\
    SWE-Pruner
    & 59.5\chg{+2.5} & 114.8\chg{+24\%} & 96.1\chg{+28\%}
    & 4.05\chg{+17\%} & 0.35\chg{+9\%}
    & 0.024\chg{0\%} & 0.052\chg{+0.052} & 4.471\chg{+18\%} \\
    \method{}
    & \textbf{60.5}\chg{+3.5} & \textbf{77.5}\chg{-16\%} & \textbf{66.2}\chg{-12\%}
    & \textbf{2.12}\chg{-39\%} & \textbf{0.24}\chg{-25\%} & \textbf{0.020}\chg{-17\%}
    & 0.044\chg{+0.044} & \textbf{2.428}\chg{-36\%} \\
    \midrule
    \rowcolor{tablemodelgreen}
    \multicolumn{9}{c}{\textbf{Deepseek-v4-Pro}} \\
    \midrule
    Vanilla
    & \underline{76.5} & \textbf{37.7} & \textbf{33.6}
    & \underline{1.03} & 0.025 & 0.013
    & \textbf{0.000} & \underline{1.068} \\
    LLMLingua-2
    & 73.5\chg{-3.0} & 54.5\chg{+45\%} & 49.9\chg{+49\%}
    & 1.70\chg{+65\%} & 0.026\chg{+4\%}
    & 0.017\chg{+31\%} & 0.034\chg{+0.034} & 1.775\chg{+66\%} \\
    LongCodeZip
    & 74.5\chg{-2.0} & 40.9\chg{+8\%} & 36.8\chg{+10\%}
    & 1.16\chg{+13\%} & \underline{0.021}\chg{-16\%} & 0.015\chg{+15\%}
    & \underline{0.016}\chg{+0.016} & 1.209\chg{+13\%} \\
    SWE-Pruner
    & \textbf{77.0}\chg{+0.5} & 44.1\chg{+17\%} & 39.1\chg{+16\%} & 1.20\chg{+17\%}
    & 0.015\chg{-40\%} & 0.015\chg{+15\%} & 0.021\chg{+0.021} & 1.253\chg{+17\%} \\
    \method{}
    & 75.0\chg{-1.5} & \underline{38.1}\chg{+1\%} & \underline{36.6}\chg{+9\%}
    & \textbf{0.82}\chg{-20\%} & 0.015\chg{-40\%} & 0.013\chg{0\%}
    & 0.021\chg{+0.021} & \textbf{0.868}\chg{-19\%} \\
    \bottomrule
    \end{tabular}
    \end{adjustbox}
    \vspace{-8pt}

    % \begin{minipage}{\textwidth}
    % \scriptsize
    % \emph{Note.} Superscripts report changes relative to Vanilla under the same agentic model.
    % For \passatone{}, changes are percentage points; for efficiency metrics except Comp, changes are relative percentages; for Comp, changes are absolute increases in million tokens.
    % \end{minipage}
    % \endgroup
\end{table*}

\subsection{RQ2: Can \method{} achieve a better efficiency-effectiveness trade-off than existing observation compression methods?}
In this RQ, we compare \method{} with existing observation compression methods to evaluate whether \method{} achieves a better efficiency-effectiveness trade-off.

\vspace{1pt} \bi{Setting.}
We compare \method{} with Vanilla and three observation compression baselines: LLMLingua-2, LongCodeZip, and SWE-Pruner.
Due to the high cost of Gemini3-Flash, we choose to evaluate all methods on Qwen3.5-35B-A3B and Deepseek-v4-Pro.
For each agentic model, all methods are evaluated on the same SWE-bench Verified instances.

\begin{figure}[t]
    \centering
    \includegraphics[width=0.9\columnwidth]{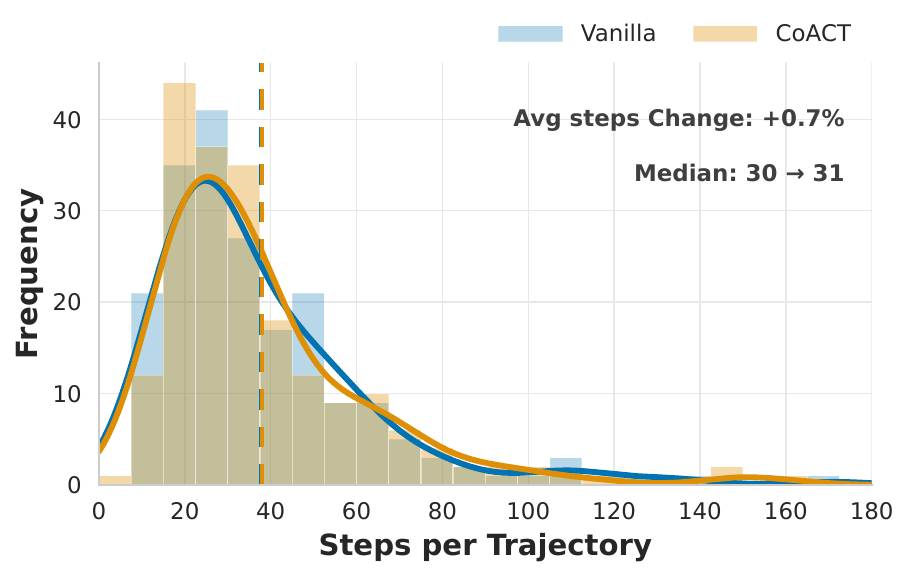}
    \caption{
    Trajectory-step distributions of Vanilla and \method{} under Deepseek-v4-Pro.
    Bars show frequency, and curves show smoothed distributions.
    }
    \label{fig:rq2_steps_distribution}
    \vspace{-10pt}
\end{figure}
    
\vspace{1pt} \bi{Results.}
\Cref{tab:rq2_observation_baselines} reports the effectiveness and efficiency metrics of \method{} and the observation compression baselines under the two agentic models.

\textbf{Under our evaluation, \method{} achieves the best efficiency-effectiveness trade-off among observation compression methods.}
Existing baselines tend to improve one side of the trade-off at the cost of the other.
For example, on Deepseek-v4-Pro, SWE-Pruner preserves \passatone{} well, but increases total token consumption by 17\%.
LongCodeZip reduces uncached input tokens, but this comes with a lower \passatone{}.
In contrast, across all evaluated agentic models, \method{} reduces total token consumption while keeping \passatone{} close to Vanilla, in line with \Cref{eq:intro_objective}.

\textbf{\method{} reduces total token consumption without increasing interaction steps.}
Observation compression would be less useful if compression forced the agent to issue more recovery actions.
This issue appears in several baselines.
For example, LLMLingua-2 increases Steps by 111\% on Qwen3.5-35B-A3B and by 45\% on Deepseek-v4-Pro.
SWE-Pruner also increases Steps by 24\% and 17\% on the two models, respectively.
We attribute these increases to useful information in observations being inadvertently removed by existing methods, which can force the agent to spend extra steps recovering missing context.
In contrast, \method{} does not substantially increase Steps on either evaluated model.
To examine more fine-grained distributional changes, we further compare the per-instance trajectory-step distributions of Vanilla and \method{} under Deepseek-v4-Pro.
As shown in \Cref{fig:rq2_steps_distribution}, the distribution of \method{} remains close to Vanilla.
This result indicates that compressing individual observations with \method{} does not come at the cost of increasing interaction steps.
    
\begin{rqanswer}
\method{} achieves the strongest efficiency-effectiveness trade-off, and compressing individual observations with \method{} does not come at the cost of increasing interaction steps.
\end{rqanswer}

\subsection{RQ3: Can \method{} complement trajectory compression for further cost reduction?}
This RQ examines whether \method{} can be combined with trajectory compression for additional savings.

\vspace{1pt} \bi{Setting.}
We combine \method{} with two representative trajectory compression methods, Sliding Window and AgentDiet.
For each combined method, \method{} first compresses the raw observation before it is appended to the trajectory.
Sliding Window or AgentDiet then compresses the accumulated trajectory using its own strategy.

We conduct experiments on Qwen3.5-35B-A3B using the same SWE-bench Verified instances as RQ1 and RQ2.
We report \passatone{}, Steps, and Tokens.
Because trajectory compression can disrupt valuable KV-cache reuse, fewer tokens do not necessarily imply lower cost.
Therefore, we also report the estimated total cost (\$) calculated according to the official API pricing~\cite{deepseek_pricing_2026, claude_pricing, openai_pricing}, where uncached input tokens are about 10x more expensive than cached input tokens.

\vspace{1pt} \bi{Results.}
\Cref{tab:rq3_trajectory_combination} reports the results of combining \method{} with trajectory compression methods.

\textbf{Trajectory compression can reduce total token consumption without lowering the final cost.}
Both trajectory compressors substantially reduce the total token consumption: Sliding Window and AgentDiet reduce total token consumption by 40.3\% and 43.7\% relative to Vanilla, respectively.
However, lower total token consumption does not translate into comparable cost savings.
Sliding Window lowers total cost by only 8.6\%, while AgentDiet even increases total cost by 10.6\%.
The key reason is that total cost depends not only on total token consumption, but also on how much of each invocation can benefit from KV-cache reuse.
Trajectory compression shortens the context at the cost of disrupting valuable prefix KV-cache reuse, causing more later tokens to be billed as uncached input.

\textbf{\method{} complements trajectory compression methods by shortening observations before they enter the trajectory.}
When combined with AgentDiet, \method{} reduces total cost from \$45.65 to \$25.88 while keeping \passatone{} nearly unchanged.
When combined with Sliding Window, \method{} reduces total token consumption from 2.27M to 1.66M and improves \passatone{} from 50.0\% to 62.0\%.
These results show that \method{} can be combined with trajectory compression methods to further reduce cost.

\begin{table}[t]
\centering
\tablecaptioninline[RQ3 combination results.]{
Comparison between trajectory compression and its combination with \method{}.
Traj. Comp. denotes the trajectory compression.
\textbf{Bold} and \underline{underlined} values mark the best and second-best results in each column.
}
\label{tab:rq3_trajectory_combination}
\setlength{\tabcolsep}{1.8pt}
\begin{tabular}{@{}llcccc@{}}
\toprule
Traj. Comp. & \method{} & \passatone{} $\uparrow$ & Steps $\downarrow$ & Tokens $\downarrow$ & Cost $\downarrow$ \\
\midrule
\multirow{2}{*}{None} & w/o \method{} & 57.0 & 92.6 & 3.80 & 41.26\\
& w/ \method{} & \underline{60.5} & \textbf{77.5} & 2.43 & \underline{27.51}\\
\midrule
\multirow{2}{*}{Sliding Window} & w/o \method{} & 50.0 & 106.2 & 2.27 & 37.71\\
& w/ \method{} & \textbf{62.0} & 86.2 & \textbf{1.66} & 31.84 \\
\midrule
\multirow{2}{*}{AgentDiet} & w/o \method{} & 59.5 & \underline{77.9} & 2.14 & 45.65\\
& w/ \method{} & 60.0 & 78.5 & \underline{2.11} & \textbf{25.88}\\
\bottomrule
\end{tabular}
\end{table}

\begin{rqanswer}
\method{} can complement trajectory compression methods to further reduce cost without sacrificing effectiveness.
\end{rqanswer}

\subsection{RQ4: How do the two rewards contribute to the trained compressor?}
In this RQ, we examine whether the two rewards used in compressor training are both necessary for achieving the final compression performance.

\vspace{1pt} \bi{Setting.}
We conduct this ablation on Qwen3.5-35B-A3B using the same SWE-bench Verified instances and mini-swe-agent scaffold as the previous RQs.
We compare \method{} with three variants of \method{}.
\ding{182} No Reward trains the same compressor with the same teacher-generated candidate compressions, but chooses among the generated compressions randomly rather than using reward guidance.
\ding{183} w/o AP removes the action-preservation reward.
\ding{184} w/o LR removes the length-reduction reward.
For fairness, all three variants use the same number of training examples as \method{}.

\begin{table}[t]
\centering
\tablecaptioninline[RQ4 reward ablation results.]{
Comparison among Vanilla, \method{}, and reward-ablation variants.
AP and LR denote the action-preservation and length-reduction rewards.
No Reward denotes the variant trained without reward guidance.
}
\label{tab:rq4_reward_ablation}
\begin{tabular}{@{}lcccc@{}}
\toprule
Variant & \passatone{} $\uparrow$ & Steps $\downarrow$ & S-Steps $\downarrow$ & Tokens $\downarrow$ \\
\midrule
Vanilla & 57.0 & 92.6 & 75.0 & 3.795\\
\midrule
No Reward & 58.0 & 81.4 & 74.3 & 2.793\\
w/o AP & 50.0 & 80.5 & 62.0 & 2.607\\
w/o LR & 61.0 & 85.5 & 68.1 & 2.930\\
\method{} & 60.5 & 77.5 & 66.2 & 2.428\\
\bottomrule
\end{tabular}
\vspace{-10pt}
\end{table}

\vspace{1pt} \bi{Results.}
\Cref{tab:rq4_reward_ablation} reports the effectiveness and efficiency metrics of all methods compared in this RQ.

\textbf{The action-preservation reward is critical for task success.}
Removing AP causes the largest performance drop, reducing \passatone{} from 60.5\% to 50.0\%.
Without AP, candidate selection is mainly driven by length reduction, so shorter compressions may be chosen even when they hurt task-solving behavior.
This confirms that AP is necessary for modeling the effectiveness constraint through next-action preservation.

\textbf{The length-reduction reward is critical for execution efficiency.}
Removing LR achieves a similar \passatone{} to \method{} (61.0\% vs. 60.5\%), but increases total token consumption from 2.428M to 2.930M.
This shows that LR is needed to further reduce observation length and improve execution efficiency.

\textbf{Combining the two rewards improves both effectiveness and efficiency.}
Without AP, the compressor reduces tokens but loses task effectiveness; without LR, it preserves task effectiveness but leaves more tokens in the trajectory.
With both rewards, \method{} keeps \passatone{} close to w/o LR (60.5\% vs. 61.0\%) while reducing total token consumption to 2.428M, the lowest among all variants.
This shows that AP and LR are complementary: AP enforces the effectiveness constraint, while LR improves efficiency within this constraint.

\begin{rqanswer}
The action-preservation reward protects task success, the length-reduction reward reduces token consumption, and combining them gives the best efficiency-effectiveness trade-off.
\end{rqanswer}

\subsection{RQ5: Does online compressor alignment improve the compressor beyond offline bootstrap?}
This RQ evaluates whether online alignment improves the compressor after offline bootstrap.

\vspace{1pt} \bi{Setting.}
We compare the offline-only compressor $c_{\phi_1}$ with the online-aligned compressor $c_{\phi_2}$ under the setup described in \Cref{sec:eval_setup}.
The offline-only compressor is trained only on the offline training set $\mathcal{D}_1$.
The online-aligned compressor continues training $c_{\phi_1}$ on the online training set $\mathcal{D}_2$.

\begin{table}[t]
\centering
\tablecaptioninline[RQ5 offline bootstrap and online alignment results.]{
RQ5 offline bootstrap and online alignment results.
}
\label{tab:rq5_on_policy}
\setlength{\tabcolsep}{3pt}
\begin{tabular}{@{}lcccc@{}}
\toprule
Compressor & \passatone{} $\uparrow$ & Steps $\downarrow$ & S-Steps $\downarrow$ & Tokens $\downarrow$ \\
\midrule
None & 57.0 & 92.6 & 75.0 & 3.795\\
Offline-only compressor & 56.0 & 78.1 & 62.9 & 2.409\\
Online-aligned compressor & 60.5 & 77.5 & 66.2 & 2.428 \\
\bottomrule
\end{tabular}
\vspace{-10pt}
\end{table}

\vspace{1pt} \bi{Results.}
\Cref{tab:rq5_on_policy} reports the effectiveness and efficiency metrics of the offline-only and online-aligned compressors.

\textbf{Online alignment improves effectiveness without sacrificing efficiency.}
As shown in \Cref{tab:rq5_on_policy}, compared with the offline-only compressor, the online-aligned compressor improves \passatone{} from 56.0\% to 60.5\%, while keeping Steps nearly unchanged and increasing total token consumption only slightly from 2.409M to 2.428M.
We attribute this effectiveness gain to the reduced mismatch between supervision construction and deployment.
Offline bootstrap constructs supervision from uncompressed trajectories, whereas deployment uses trajectories where earlier observations may already be compressed.
Online alignment constructs supervision from compressed trajectories, making it closer to deployment.

\begin{rqanswer}
Online alignment reduces the mismatch between training and deployment, improving effectiveness without sacrificing efficiency.
\end{rqanswer}

\section{Discussion}
\label{sec:discussion}

\begin{table}[ht]
\centering
\tablecaptioninline[Generalization across teacher models.]{
Generalization across teacher models.
}
\label{tab:teacher_model_effect}
\begin{tabular}{@{}lcccc@{}}
\toprule
Teacher & \passatone{} $\uparrow$ & Steps $\downarrow$ & S-Steps $\downarrow$ & Tokens $\downarrow$ \\
\midrule
Gemini3-Flash & 60.5 & 77.5 & 66.2 & 2.428 \\
Deepseek-v4-Pro & 52.5 & 76.5 & 60.5 & 2.339 \\
\bottomrule
\end{tabular}
\vspace{-10pt}
\end{table}

\subsection{Generalization Across Teacher Models}
\label{sec:discussion_teacher_model}

During reward-guided supervision construction, \method{} uses a teacher model to generate candidate compressions for each raw observation.
To examine whether the trained observation compressor can generalize across different teacher models, we use Qwen3.5-35B-A3B as the agentic model and evaluate on the same SWE-bench Verified instances used in the main experiments.
We fix the compressor initialization and evaluation setting, and vary only the teacher model used to construct the supervision data.
As shown in \Cref{tab:teacher_model_effect}, both teacher models lead to compressors with lower total token consumption than Vanilla under Qwen3.5-35B-A3B.
The Gemini3-Flash teacher preserves task success better in this setting, achieving 60.5\% \passatone{} compared with 52.5\% for Deepseek-v4-Pro.
We therefore use Gemini3-Flash as the default teacher in the main experiments.

\begin{table}[ht]
\centering
\tablecaptioninline[Generalization to unseen agentic models.]{
Generalization to unseen agentic models.
The second column shows which agentic model generated the trajectories used for compressor training.
}
\label{tab:transfer_models}
\setlength{\tabcolsep}{3pt}
\begin{tabular}{@{}llcc@{}}
\toprule
Evaluation Model & Compressor Trained With & \passatone{} $\uparrow$ & Tokens $\downarrow$ \\
\midrule
\multirow{2}{*}{Qwen3.5-35B-A3B}
& Qwen3.5-35B-A3B & 60.5 & 2.428 \\
& Deepseek-v4-Pro & 60.0 & 2.518 \\
\midrule
\multirow{2}{*}{Deepseek-v4-Pro}
& Qwen3.5-35B-A3B & 74.5 & 0.863 \\
& Deepseek-v4-Pro & 75.0 & 0.868 \\
\bottomrule
\end{tabular}
\vspace{-10pt}
\end{table}

\subsection{Generalization to Unseen Agentic Models}
\label{sec:discussion_transfer}

For a new agentic model, re-collecting trajectories with that model would add deployment cost.
We therefore test whether a compressor trained on trajectories sampled by one agentic model can work for another: we train one compressor on trajectories sampled by Qwen3.5-35B-A3B and another on trajectories sampled by Deepseek-v4-Pro, then evaluate each compressor with both agentic models on the same SWE-bench Verified instances.

\Cref{tab:transfer_models} shows that using trajectories sampled by another agentic model for compressor training gives results close to using trajectories sampled by the evaluation model itself.
For example, with Deepseek-v4-Pro as the evaluation model, the compressor trained on trajectories sampled by Qwen3.5-35B-A3B obtains 74.5\% \passatone{} and 0.863M tokens, compared with 75.0\% and 0.868M for the compressor trained on Deepseek-v4-Pro trajectories.
These small gaps suggest that an agentic model can use an observation compressor trained on trajectories sampled by another agentic model, without re-sampling trajectories with its own model.

\subsection{Reliability of the Action-Similarity Function}
\label{sec:discussion_action_similarity}

We validate whether the action-similarity function used in the action-preservation reward matches human judgments about whether two actions are the same.
We randomly sample 100 action pairs from training data used in our main evaluation and compute their action-similarity scores.
Following the same threshold used to accept candidate compressions, we treat a pair as the same action if its score is at least $\theta=0.6$, and as different otherwise.
Human annotators provide only binary labels for each pair, indicating whether the two actions are the same.
The resulting decisions agree with human labels on 80.0\% of the pairs, supporting its use for candidate filtering.

\subsection{Threats to Validity}
\label{sec:threats}

\ding{182} \bi{Generalization Across Agent Settings.}
A potential concern is whether \method{} depends on a specific agent setting, since coding-agent workflows can vary in their prompts, tools, and action formats.
We mitigate this concern by compressing only observations, leaving the agent policy, tool interface, and execution loop unchanged.
We also evaluate three agentic models and study cross-model transfer in \Cref{sec:discussion_transfer}, supporting broader applicability.

\ding{183} \bi{Reliability of NAP.}
A potential concern is whether next-action preservation (NAP) is a reliable proxy for \passatone{}.
We address this concern from both theoretical and empirical perspectives.
At the theoretical level, if each compression preserves the agent's next action, the compressed agent follows the same action sequence as the uncompressed agent and therefore obtains the same \passatone{}.
At the practical level, the main experiments show that using NAP to guide compression reduces token consumption while keeping \passatone{} close to the uncompressed agent, and \Cref{tab:rq4_reward_ablation} further shows that removing the NAP-based action-preservation reward weakens the efficiency-effectiveness trade-off.

\ding{184} \bi{Coverage of Comparison Methods.}
Our baseline set may not sufficiently cover the relevant compression methods for coding agents.
We mitigate this threat by choosing representative methods spanning general text compression, code-aware compression, coding-agent-specific observation compression.
We further include Sliding Window and AgentDiet to evaluate whether \method{} complements widely used trajectory-compression methods.
This allows us to compare \method{} against both observation-level and trajectory-level alternatives.

\section{Conclusion}
\label{sec:conclusion}

In this paper, we propose \method{}, an action-preserving observation compression method for coding agents.
Through comprehensive experiments on SWE-bench Verified with three agentic models, we show that \method{} consistently reduces total token consumption while preserving task-solving effectiveness.
We further demonstrate that \method{} achieves a better efficiency-effectiveness trade-off than existing observation-compression baselines and can complement trajectory compression for additional cost reduction.
These results highlight action-preserving observation compression as a promising direction for reducing coding-agent inference cost while maintaining effectiveness.

\section{Data Availability}
\ifsubmission
The source code and related datasets can be accessed at: \url{https://anonymous.4open.science/r/CoACT-8BFC/}.
\else
The source code and related datasets can be accessed at: \url{https://github.com/TsinghuaISE/CoACT}.
\fi

\bibliographystyle{IEEEtran}
\bibliography{references}

@article{yang2024swe,
  title={Swe-agent: Agent-computer interfaces enable automated software engineering},
  author={Yang, John and Jimenez, Carlos E and Wettig, Alexander and Lieret, Kilian and Yao, Shunyu and Narasimhan, Karthik and Press, Ofir},
  journal={Advances in Neural Information Processing Systems},
  volume={37},
  pages={50528--50652},
  year={2024}
}

@inproceedings{wang2025openhands,
  title={Openhands: An open platform for ai software developers as generalist agents},
  author={Wang, Xingyao and Li, Boxuan and Song, Yufan and Xu, Frank F and Tang, Xiangru and Zhuge, Mingchen and Pan, Jiayi and Song, Yueqi and Li, Bowen and Singh, Jaskirat and others},
  booktitle={International Conference on Learning Representations},
  volume={2025},
  pages={65882--65919},
  year={2025}
}

@article{xia2024agentless,
  title={Agentless: Demystifying llm-based software engineering agents},
  author={Xia, Chunqiu Steven and Deng, Yinlin and Dunn, Soren and Zhang, Lingming},
  journal={arXiv preprint arXiv:2407.01489},
  year={2024}
}

@article{gao2025trae,
  title={Trae agent: An llm-based agent for software engineering with test-time scaling},
  author={Gao, Pengfei and Tian, Zhao and Meng, Xiangxin and Wang, Xinchen and Hu, Ruida and Xiao, Yuanan and Liu, Yizhou and Zhang, Zhao and Chen, Junjie and Gao, Cuiyun and others},
  journal={arXiv preprint arXiv:2507.23370},
  year={2025}
}

@article{xiao2025reducing,
  title={Reducing Cost of LLM Agents with Trajectory Reduction},
  author={Xiao, Yuan-An and Gao, Pengfei and Peng, Chao and Xiong, Yingfei},
  journal={arXiv preprint arXiv:2509.23586},
  year={2025}
}

@article{wang2026swe,
  title={SWE-Pruner: Self-Adaptive Context Pruning for Coding Agents},
  author={Wang, Yuhang and Shi, Yuling and Yang, Mo and Zhang, Rongrui and He, Shilin and Lian, Heng and Chen, Yuting and Ye, Siyu and Cai, Kai and Gu, Xiaodong},
  journal={arXiv preprint arXiv:2601.16746},
  year={2026}
}

@article{lindenbauer2025complexity,
  title={The Complexity Trap: Simple Observation Masking Is as Efficient as LLM Summarization for Agent Context Management},
  author={Lindenbauer, Tobias and Slinko, Igor and Felder, Ludwig and Bogomolov, Egor and Zharov, Yaroslav},
  journal={arXiv preprint arXiv:2508.21433},
  year={2025}
}

@article{fan2025swe,
  title={Swe-effi: Re-evaluating software ai agent system effectiveness under resource constraints},
  author={Fan, Zhiyu and Vasilevski, Kirill and Lin, Dayi and Chen, Boyuan and Chen, Yihao and Zhong, Zhiqing and Zhang, Jie M and He, Pinjia and Hassan, Ahmed E},
  journal={arXiv preprint arXiv:2509.09853},
  year={2025}
}

@inproceedings{pan2024llmlingua,
  title={Llmlingua-2: Data distillation for efficient and faithful task-agnostic prompt compression},
  author={Pan, Zhuoshi and Wu, Qianhui and Jiang, Huiqiang and Xia, Menglin and Luo, Xufang and Zhang, Jue and Lin, Qingwei and R{\"u}hle, Victor and Yang, Yuqing and Lin, Chin-Yew and others},
  booktitle={Findings of the Association for Computational Linguistics: ACL 2024},
  pages={963--981},
  year={2024}
}

@inproceedings{jimenez2024swe,
  title={Swe-bench: Can language models resolve real-world github issues?},
  author={Jimenez, Carlos E and Yang, John and Wettig, Alexander and Yao, Shunyu and Pei, Kexin and Press, Ofir and Narasimhan, Karthik},
  booktitle={International Conference on Learning Representations},
  volume={2024},
  pages={54107--54157},
  year={2024}
}

@article{merrill2026terminal,
  title={Terminal-bench: Benchmarking agents on hard, realistic tasks in command line interfaces},
  author={Merrill, Mike A and Shaw, Alexander G and Carlini, Nicholas and Li, Boxuan and Raj, Harsh and Bercovich, Ivan and Shi, Lin and Shin, Jeong Yeon and Walshe, Thomas and Buchanan, E Kelly and others},
  journal={arXiv preprint arXiv:2601.11868},
  year={2026}
}

@misc{stackoverflow2025,
	title = {{AI} {\textbar} 2025 {Stack} {Overflow} {Developer} {Survey}},
	url = {https://survey.stackoverflow.co/2025/ai},
	language = {en},
	urldate = {2026-05-19},
}

@misc{cursorCursor,
	author = {},
	title = {{C}ursor --- cursor.com},
	howpublished = {\url{https://cursor.com/}},
	year = {},
}

@misc{claude_pricing,
	title = {Pricing},
	url = {https://platform.claude.com/docs/en/about-claude/pricing},
	language = {en-US},
	urldate = {2026-05-19},
	journal = {Claude API Docs},
}

@misc{openai_pricing,
	title = {{OpenAI} {API} {Pricing}},
	url = {https://openai.com/api/pricing/},
	language = {en-US},
	urldate = {2026-05-19},
	journal = {OpenAI},
}

@inproceedings{jiang2023llmlingua,
  title={Llmlingua: Compressing prompts for accelerated inference of large language models},
  author={Jiang, Huiqiang and Wu, Qianhui and Lin, Chin-Yew and Yang, Yuqing and Qiu, Lili},
  booktitle={Proceedings of the 2023 conference on empirical methods in natural language processing},
  pages={13358--13376},
  year={2023}
}

@inproceedings{jiang2024longllmlingua,
  title={Longllmlingua: Accelerating and enhancing llms in long context scenarios via prompt compression},
  author={Jiang, Huiqiang and Wu, Qianhui and Luo, Xufang and Li, Dongsheng and Lin, Chin-Yew and Yang, Yuqing and Qiu, Lili},
  booktitle={Proceedings of the 62nd Annual Meeting of the Association for Computational Linguistics (Volume 1: Long Papers)},
  pages={1658--1677},
  year={2024}
}

@article{chirkova2025provence,
  title={Provence: efficient and robust context pruning for retrieval-augmented generation},
  author={Chirkova, Nadezhda and Formal, Thibault and Nikoulina, Vassilina and Clinchant, St{\'e}phane},
  journal={arXiv preprint arXiv:2501.16214},
  year={2025}
}

@inproceedings{liskavets2025prompt,
  title={Prompt compression with context-aware sentence encoding for fast and improved llm inference},
  author={Liskavets, Barys and Ushakov, Maxim and Roy, Shuvendu and Klibanov, Mark and Etemad, Ali and Luke, Shane K},
  booktitle={Proceedings of the AAAI Conference on Artificial Intelligence},
  volume={39},
  number={23},
  pages={24595--24604},
  year={2025}
}

@article{shi2025longcodezip,
  title={Longcodezip: Compress long context for code language models},
  author={Shi, Yuling and Qian, Yichun and Zhang, Hongyu and Shen, Beijun and Gu, Xiaodong},
  journal={arXiv preprint arXiv:2510.00446},
  year={2025}
}

@article{he2025codepromptzip,
  title={Codepromptzip: Code-specific prompt compression for retrieval-augmented generation in coding tasks with lms},
  author={He, Pengfei and Wang, Shaowei and Chen, Tse-Hsun},
  journal={arXiv preprint arXiv:2502.14925},
  year={2025}
}

@misc{claude_code,
	title = {Claude {Code} by {Anthropic} {\textbar} {AI} {Coding} {Agent}, {Terminal}, {IDE}},
	url = {https://claude.com/product/claude-code},
	language = {en},
	urldate = {2026-05-20},
}

@article{sun2025scaling,
  title={Scaling long-horizon llm agent via context-folding},
  author={Sun, Weiwei and Lu, Miao and Ling, Zhan and Liu, Kang and Yao, Xuesong and Yang, Yiming and Chen, Jiecao},
  journal={arXiv preprint arXiv:2510.11967},
  year={2025}
}

@article{ye2025agentfold,
  title={AgentFold: Long-Horizon Web Agents with Proactive Context Management},
  author={Ye, Rui and Zhang, Zhongwang and Li, Kuan and Yin, Huifeng and Tao, Zhengwei and Zhao, Yida and Su, Liangcai and Zhang, Liwen and Qiao, Zile and Wang, Xinyu and others},
  journal={arXiv preprint arXiv:2510.24699},
  year={2025}
}

@article{lu2025scaling,
  title={Scaling llm multi-turn rl with end-to-end summarization-based context management},
  author={Lu, Miao and Sun, Weiwei and Du, Weihua and Ling, Zhan and Yao, Xuesong and Liu, Kang and Chen, Jiecao},
  journal={arXiv preprint arXiv:2510.06727},
  year={2025}
}

@article{wan2025compass,
  title={Compass: Enhancing agent long-horizon reasoning with evolving context},
  author={Wan, Guangya and Ling, Mingyang and Ren, Xiaoqi and Han, Rujun and Li, Sheng and Zhang, Zizhao},
  journal={arXiv preprint arXiv:2510.08790},
  year={2025}
}

@article{kang2025acon,
  title={Acon: Optimizing context compression for long-horizon llm agents},
  author={Kang, Minki and Chen, Wei-Ning and Han, Dongge and Inan, Huseyin A and Wutschitz, Lukas and Chen, Yanzhi and Sim, Robert and Rajmohan, Saravan},
  journal={arXiv preprint arXiv:2510.00615},
  year={2025}
}

@article{ren2026self,
  title={A Self-Evolving Framework for Efficient Terminal Agents via Observational Context Compression},
  author={Ren, Jincheng and Wu, Siwei and Li, Yizhi and Zhu, Kang and Xu, Shu and Feng, Boyu and Yuan, Ruibin and Zhang, Wei and Batista-Navarro, Riza and Yang, Jian and others},
  journal={arXiv preprint arXiv:2604.19572},
  year={2026}
}

@article{feng2026agentswing,
  title={AgentSwing: Adaptive Parallel Context Management Routing for Long-Horizon Web Agents},
  author={Feng, Zhaopeng and Su, Liangcai and Zhang, Zhen and Wang, Xinyu and Zhang, Xiaotian and Wang, Xiaobin and Fang, Runnan and Zhang, Qi and Li, Baixuan and Cai, Shihao and others},
  journal={arXiv preprint arXiv:2603.27490},
  year={2026}
}

@article{yao2022react,
  title={React: Synergizing reasoning and acting in language models},
  author={Yao, Shunyu and Zhao, Jeffrey and Yu, Dian and Du, Nan and Shafran, Izhak and Narasimhan, Karthik and Cao, Yuan},
  journal={arXiv preprint arXiv:2210.03629},
  year={2022}
}

@article{yang2026swe,
  title={Swe-smith: Scaling data for software engineering agents},
  author={Yang, John and Lieret, Kilian and Jimenez, Carlos and Wettig, Alexander and Khandpur, Kabir and Zhang, Yanzhe and Hui, Binyuan and Press, Ofir and Schmidt, Ludwig and Yang, Diyi},
  journal={Advances in Neural Information Processing Systems},
  volume={38},
  year={2026}
}

@article{xu2026deepseek,
  title={DeepSeek-V4: Towards Highly Efficient Million-Token Context Intelligence},
  author={Xu, Anyi and Lin, Bangcai and Xue, Bing and Wang, Bingxuan and Xu, Bingzheng and Wu, Bochao and Zhang, Bowei and Lin, Chaofan and Dong, Chen and Ling, Chenchen and others},
  journal={arXiv preprint arXiv:2606.19348},
  year={2026}
}

@article{hu2022lora,
  title={Lora: Low-rank adaptation of large language models.},
  author={Hu, Edward J and Shen, Yelong and Wallis, Phillip and Allen-Zhu, Zeyuan and Li, Yuanzhi and Wang, Shean and Wang, Liang and Chen, Weizhu and others},
  journal={Iclr},
  volume={1},
  number={2},
  pages={3},
  year={2022}
}

@misc{qwen3.5-35b-a3b_2026,
	title = {Qwen/{Qwen3}.5-{35B}-{A3B} · {Hugging} {Face}},
	url = {https://huggingface.co/Qwen/Qwen3.5-35B-A3B},
	urldate = {2026-06-28},
	month = mar,
	year = {2026},
}

@misc{qwen3.5-4b_2026,
	title = {Qwen/{Qwen3}.5-{4B} · {Hugging} {Face}},
	url = {https://huggingface.co/Qwen/Qwen3.5-4B},
	urldate = {2026-06-28},
	month = mar,
	year = {2026},
}

@misc{deepseek_pricing_2026,
	title = {Models \& {Pricing} {\textbar} {DeepSeek} {API} {Docs}},
	url = {https://api-docs.deepseek.com/quick_start/pricing},
	language = {en},
	urldate = {2026-06-28},
}

@misc{gemini_2025,
	title = {Gemini 3 {Flash}: frontier intelligence built for speed},
	shorttitle = {Gemini 3 {Flash}},
	url = {https://blog.google/products-and-platforms/products/gemini/gemini-3-flash/},
	language = {en-us},
	urldate = {2026-06-28},
	journal = {Google},
	month = dec,
	year = {2025},
}

@article{bai2026ai,
  title={How do AI agents spend your money? Analyzing and predicting token consumption in agentic coding tasks},
  author={Bai, Longju and Huang, Zhemin and Wang, Xingyao and Sun, Jiao and Mihalcea, Rada and Brynjolfsson, Erik and Pentland, Alex and Pei, Jiaxin},
  journal={arXiv preprint arXiv:2604.22750},
  year={2026}
}

@article{gao2025more,
  title={More with less: An empirical study of turn-control strategies for efficient coding agents},
  author={Gao, Pengfei and Peng, Chao},
  journal={arXiv preprint arXiv:2510.16786},
  year={2025}
}

@inproceedings{liu2026context,
  title={Context as a tool: Context management for long-horizon swe-agents},
  author={Liu, Shukai and Jiang, Bo and Yang, Jian and Li, Yizhi and Guo, Jinyang and Liu, Xianglong and Dai, Bryan},
  booktitle={Findings of the Association for Computational Linguistics: ACL 2026},
  pages={20604--20617},
  year={2026}
}

@article{verma2026active,
  title={Active Context Compression: Autonomous Memory Management in LLM Agents},
  author={Verma, Nikhil},
  journal={arXiv preprint arXiv:2601.07190},
  year={2026}
}

@article{jia2026compressing,
  title={Compressing code context for llm-based issue resolution},
  author={Jia, Haoxiang and Barr, Earl T and Mechtaev, Sergey},
  journal={arXiv preprint arXiv:2603.28119},
  year={2026}
}

@article{cai2025ai,
  title={AI-Driven Self-Evolving Software: A Promising Path Toward Software Automation},
  author={Cai, Liyi and Ren, Yijie and Zhang, Yitong and Li, Jia},
  journal={arXiv preprint arXiv:2510.00591},
  year={2025}
}

@article{li2026papers,
  title={What Papers Don't Tell You: Recovering Tacit Knowledge for Automated Paper Reproduction},
  author={Li, Lehui and Wang, Ruining and Song, Haochen and Mao, Yaoxin and Zhang, Tong and Wang, Yuyao and Fan, Jiayi and Zhang, Yitong and Ye, Jieping and Zhang, Chengqi and others},
  journal={arXiv preprint arXiv:2603.01801},
  year={2026}
}

@misc{zhang2026environmental,
      title={Environmental Injection Attacks against GUI Agents in Realistic Dynamic Environments}, 
      author={Yitong Zhang and Ximo Li and Liyi Cai and Jia Li},
      year={2026},
      eprint={2509.11250},
      archivePrefix={arXiv},
      primaryClass={cs.CR},
      url={https://arxiv.org/abs/2509.11250}, 
}

@inproceedings{hussain2021command,
  title={Command similarity measurement using nlp},
  author={Hussain, Zafar and Nurminen, Jukka K and Mikkonen, Tommi and Kowiel, Marcin},
  booktitle={10th Symposium on Languages, Applications and Technologies (SLATE 2021)},
  pages={13--1},
  year={2021},
  organization={Schloss Dagstuhl--Leibniz-Zentrum f{\"u}r Informatik}
}

\end{document}